\documentclass[prl,twocolumn,preprintnumbers,amsmath,amssymb,floatfix]{revtex4-2}
\usepackage{graphicx}
\usepackage{epstopdf}
\usepackage{float}
\usepackage{placeins}
\usepackage{hyperref,graphicx}
\usepackage{dcolumn}
\usepackage{bm}
\usepackage{color}
\usepackage{xcolor}
\usepackage{natbib}
\usepackage{amsmath}
\usepackage{amssymb}
\usepackage{multirow} 
\usepackage{hyperref}
\usepackage{stackengine}
\hypersetup{colorlinks=true, linkcolor=blue, citecolor=red, urlcolor=magenta, pdftitle={BFO-CFO-SL}, pdfauthor={Monirul Shaikh}}
\begin{document}
\begin{center}
\begin{large}
\end{large}
\end{center}

\title{Polar Charge-Ordered States in BiFeO$_3$/CaFeO$_3$ Superlattice}
\author{Rajan Gowsalya$^\P$}
\author{Monirul Shaikh$^{\P, \ddag}$}
\email{msk.phe@gmail.com}
\author{Sathiyamoorthy Buvaneswaran$^{\P}$}
\author{Saurabh Ghosh$^{\P, \ddag}$}
\email{saurabhghosh2802@gmail.com}
\affiliation{$^\P$Department of Physics and Nanotechnology, SRM Institute of Science and Technology, Kattankulathur - 603 203, Tamil Nadu, India}
\affiliation{$^{\ddag}$Center for Advanced Computational and Theoretical Sciences, SRM Institute of Science and Technology,
Kattankulathur - 603 203, Tamil Nadu, India}

\date{\today}
\begin{abstract}
Oxide superlattices represent a potent avenue for tailoring emergent electronic phases through sophisticated interfacial charge transfer and dynamic lattice distortions. This study systematically investigates the structural and electronic attributes of the BiFeO$_3$/CaFeO$_3$ superlattice, leveraging a comprehensive approach that integrates first-principles computations with detailed symmetry-mode analysis. The strategic integration of polar bismuth ferrite alongside charge-transfer calcium ferrite instigates profound lattice instabilities, notably manifest in octahedral rotations and cooperative FeO$_6$ breathing modes that might not necessarily be soft. However, their synergistic coupling stabilizes a non-centrosymmetric $Pc$ ground state that intrinsically features polar charge ordering of Fe ions. This resultant phase ingeniously unifies C-type antiferromagnetism with robust ferroelectric semiconductor characteristics, exhibiting a calculated indirect band gap of about 0.6 eV. Our discoveries firmly establish ferrite superlattices as an exceptionally versatile and tunable platform for the rational design of next-generation multifunctional materials, offering precise control over polarization, charge ordering phenomena, and electronic transport behavior via advanced interface and strain engineering techniques.
\end{abstract}
\maketitle
\section{\label{sec:level1}I. INTRODUCTION}
Artificial oxide superlattices have emerged as an important platform for discovering emergent phases that arise from the interplay between lattice, charge, spin, and orbital degrees of freedom \cite{mannhart2010oxide,hwang2012emergent,ohtomo2004high,ramesh2019creating,chakhalian2012whither}. The development of atomic-layer growth techniques such as molecular beam epitaxy (MBE) and pulsed laser deposition (PLD) now enables the fabrication of transition-metal oxide heterostructures with unit-cell precision \cite{schlom2008thin,haeni2004room,lee2010strong,lei2017constructing}. In these artificially layered systems, broken symmetry at interfaces can generate structural and electronic states that do not occur in the corresponding bulk compounds. As a consequence, oxide superlattices provide a versatile route for designing functional materials in which lattice distortions, electronic structure, and magnetic order can be engineered through interface control \cite{rondinelli2011structure, dominguez2020length}.

Perovskite oxides are particularly suitable for such investigations because their corner-sharing BO$_6$ octahedral framework supports a wide range of cooperative lattice distortions. Octahedral rotations, tilts, breathing distortions, and polar displacements can interact strongly through symmetry-allowed coupling terms, giving rise to complex structural phase behavior \cite{lee2010strong,rondinelli2011structure, stroppa2013hybrid}. The cooperative interaction among these structural modes can stabilize new crystal structures and substantially modify the electronic properties of the material. In layered oxide heterostructures, such coupled distortions can break inversion symmetry and generate polar states even in systems that are not intrinsically ferroelectric \cite{benedek2011hybrid, Rondinelli2010substrate, benedek2013there, Ghosh2015Linear, Shaikh2020Strain}. Understanding how these lattice instabilities interact in artificial structures is therefore essential for designing oxide heterostructures with tailored electronic functionality.

Ferrite perovskites provide a particularly rich playground in this context due to the strong coupling between lattice distortions and electronic degrees of freedom. Among them, CaFeO$_3$ is a prototypical negative charge-transfer oxide that undergoes a metal–insulator transition accompanied by charge disproportionation and breathing distortions of the FeO$_6$ octahedra \cite{Rogge2018negative, Leonov2022negative}. In the insulating phase, the Fe ions become inequivalent, forming an ordered arrangement that is closely tied to the structural breathing distortion of the octahedral network \cite{Zhang2017BD}. Because the electronic ground state in CaFeO$_3$ is highly sensitive to lattice distortions, this material has gained considerable attention as a building block for oxide heterostructures in which electronic phases can be manipulated through structural engineering at interfaces \cite{Rogge2018negative}.

In contrast, BiFeO$_3$ represents one of the most extensively studied multiferroic perovskites \cite{ramesh2007multiferroics, chu2008electric, seidel2009conduction, lebeugle2007very, tokunaga2015magnetic}, exhibiting both robust antiferromagnetic order and large spontaneous ferroelectric polarization at room temperature. The strong polarization originates from the stereochemically active Bi$^{3+}$ lone pair, which drives large off-center displacements of the Bi ions and produces substantial polar distortions of the lattice \cite{wu2012effects, noguchi2022origin, kuo2016single, Ederer2005weak, chen2018complex}. Owing to this intrinsic polarity, BiFeO$_3$-based heterostructures have attracted significant attention as a route for controlling interfacial electronic and magnetic properties through ferroelectric polarization \cite{yao2025control}.

Combining BiFeO$_3$ with charge-transfer ferrites therefore offers an appealing strategy for engineering new interfacial electronic phases \cite{hwang2012emergent, mannhart2010oxide, chakhalian2014Colloquium, rondinelli2011structure}. In particular, a superlattice composed of BiFeO$_3$ and CaFeO$_3$ introduces an interface between Fe$^{3+}$ and Fe$^{4+}$ ions, which naturally generates charge imbalance across the heterostructure. Such interfacial charge redistribution can couple strongly to structural distortions of the FeO$_6$ octahedra, including breathing modes and rotational distortions. The interplay between these lattice distortions and charge transfer may stabilize charge-ordered states that simultaneously break inversion symmetry, leading to polar charge-ordered phases. From an experimental perspective, such states are especially attractive because they could enable electric-field control of electronic properties in oxide heterostructures \cite{heron2014deterministic, garcia2009giant}.

Recent experimental and theoretical studies have demonstrated that ferrite-based superlattices can host rich structural and electronic phase behavior arising from the interplay between octahedral distortions and charge ordering \cite{zhou2025ferromagnetism, chakhalian2014Colloquium, Lopez2018Spin, rout2023ferromagnetism}. Layered perovskite architectures are known to stabilize lattice instabilities that are suppressed in bulk compounds, thereby opening new pathways for tuning electronic phases through interface design. In our recent research on LaFeO$_3$/CaFeO$_3$ superlattice \cite{shaikh2026polar}, we showed that cooperative coupling among multiple lattice distortions can drive an unconventional insulator–metal–insulator transition. This study highlighted the importance of multimode structural coupling in ferrite heterostructures and demonstrated how lattice instabilities can strongly influence the electronic ground state.

Motivated by these insights, the present work explores a different route toward electronic phase control by incorporating a strongly polar component into the ferrite superlattice. Specifically, we investigate the structural and electronic properties of a BiFeO$_3$/CaFeO$_3$ superlattice using first-principles density functional theory \cite{hohenberg} combined with symmetry-mode analysis. In this system, the coexistence of Bi-driven polar distortions and interfacial charge imbalance provides a favorable environment for the emergence of polar charge ordering. 

Starting from a high-symmetry $P4/mmm$ reference structure, we identify lattice instabilities associated with octahedral rotations of FeO$_6$ octahedra and A-site antiferroelectric displacement between two successive layer. Our analysis reveals that the cooperative interaction between these structural modes together with breathing distortions of FeO$_6$ octahedra stabilizes a non-centrosymmetric $Pc$ ground state characterized by polar charge ordering of the Fe ions. However, breathing distortions of FeO$_6$ octahedra are found to be hard modes. The resulting phase exhibits C-type antiferromagnetic ordering together with a semiconducting electronic structure. These results demonstrate that combining a strongly polar oxide with a charge-transfer ferrite can generate polar charge-ordered states and tunable electronic phases in artificial heterostructures. This work highlights the potential of ferrite superlattices as a promising platform for designing multifunctional materials in which polarization, charge ordering, and electronic structure can be controlled through interface engineering.
\section{II. METHODOLOGY}
Density functional theory (DFT) calculations \cite{hohenberg} are performed to compute total energies and optimized geometries of various superlattices studied in this work using Vienna \textit{ab-initio} simulation package (VASP) \cite{kresseplanewave}. The generalized gradient approximation (GGA) \cite{wahlPAW} is considered to model the variation in electron density, whereas projector augmented wave (PAW) potentials are used for electron-ion interaction in the many-body Hamiltonian \cite{joubertPAW}. The Perdew-Burke-Ernzerhof revision for solids (PBEsol) functional \cite{PerdewGGA} is considered for the exchange-correlation part. Local spin density approximation (LSDA) + $U$ \cite{LSDA+U} method is considered throughout our calculations to integrate the static $d-d$ Coulomb interactions. To draw and analyze the geometry of our three-dimensional systems, we implement the visualization for electronic and structural analysis (VESTA) software \cite{vesta}.

In all calculations, we have considered a 500 eV  PAW energy cut-off with a $\Gamma$-centered 6 $\times$ 6 $\times$ 4 k$-$mesh for reciprocal lattice projection. We have chosen the on-site $d-d$ Coulomb interaction parameter $U_{eff}$ (= $U-J$) to be 5.0 eV, which satisfies the charge state of Fe for both BiFeO$_3$ and CaFeO$_3$. Further, the electronic and magnetic properties match well with the orthorhombic bulk BiFeO$_3$ \cite{BFO}. and CaFeO$_3$ \cite{shaikh2026polar}. Hence, we consider $U_{eff}$ = 5.0 eV for the rest of our calculations. The total energy and Hellman-Feynman force converge unless they reach the criteria 10$^{-5}$ eV and 10$^{-2}$ eV/\AA, respectively. To achieve insight into electron-phonon coupling, we perform phonon calculations on the undistorted $P4/mmm$ phase using the finite difference method as implemented in VASP. To investigate the lattice dynamics, each normalized eigenvector is `frozen' into the undistorted structure \cite{frozen-phonon}.
\section{III. RESULTS AND DISCUSSIONS}
\begin{figure*}
\centering
\includegraphics[width=0.8\linewidth]{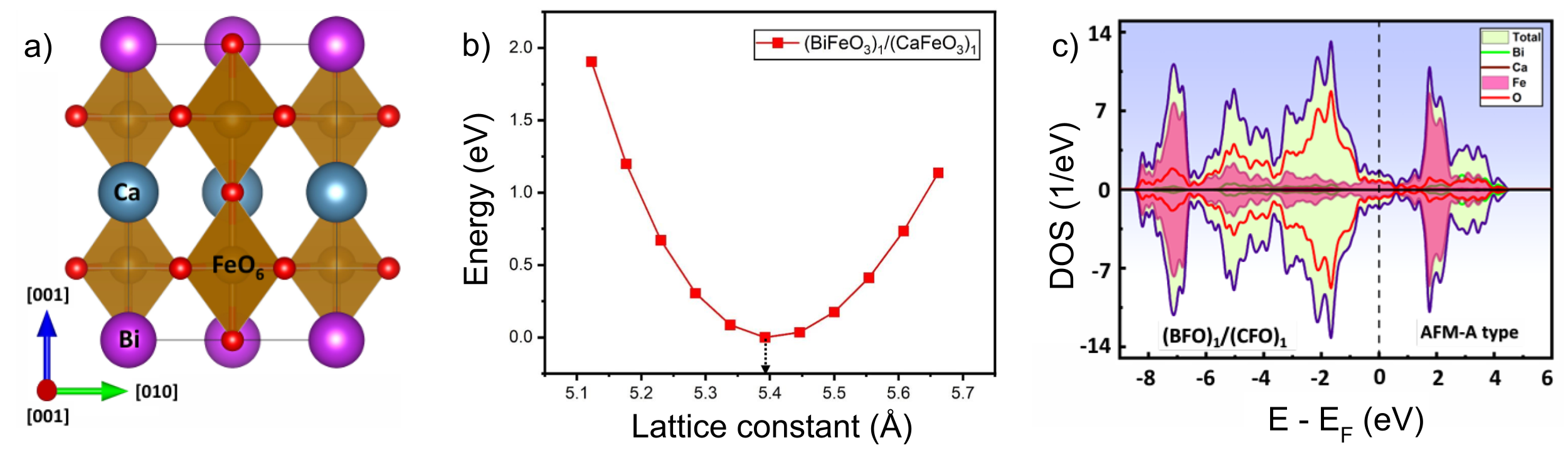}\vspace{-0pt}
\caption {(a) High-symmetry $P4/mmm$ phase of BiFeO$_3$/CaFeO$_3$ superlattice. Crystal structure with Bi (violet), Ca (green), Fe (golden color), and O (red) atoms. (b) Energy versus in-plane lattice parameter optimization for the $P4/mmm$ phase, showing a minimum at $a = 5.39 \AA$. (c) Partial density of states of high-symmetry phase showing metallic character with strong $Fe-3d$ and $O-2p$ hybridization near E$_F$}
\label{Figure1}
\end{figure*}
\subsection{A. Significance of the high-symmetry reference phase}
Figure \ref{Figure1}a shows the crystal structure of the BiFeO$_3$/CaFeO$_3$ superlattice in the ideal tetragonal $P4/mmm$ symmetry. Structurally, the $P4/mmm$ phase represents an undistorted perovskite reference structure in which the FeO$_6$ octahedra are perfectly aligned without rotations, tilts, or breathing distortions. In Glazer notation, this corresponds to the configuration $a^0a^0c^0$ \cite{Glazer1972, Glazer1975}. The centrosymmetric nature of this phase makes it an appropriate starting point for identifying symmetry-breaking lattice instabilities that can lower the total energy and generate polar or charge-ordered states. 

To determine the equilibrium lattice parameters of the high-symmetry phase, the total energy of the superlattice was calculated as a function of the in-plane lattice constant. The resulting energy-versus-lattice-parameter curve is presented in Figure \ref{Figure1}b. The calculations reveal a well-defined minimum in the total energy at an in-plane lattice parameter of approximately $a=5.39 \AA$, indicating the optimized structural configuration of the $P4/mmm$ reference phase. This optimized lattice parameter reflects the compromise between the structural preferences of the two constituent materials. BiFeO$_3$ tends to favor polar distortions driven by the stereochemically active Bi$^{3+}$ lone pair, whereas CaFeO$_3$ is prone to breathing distortions associated with charge disproportionation of Fe ions. The equilibrium lattice constant, therefore, represents the structural environment in which these competing tendencies coexist before any symmetry-breaking distortions are introduced.

The electronic properties of the optimized $P4/mmm$ structure are examined through the partial density of states (PDOS) shown in Figure \ref{Figure1}c. The PDOS reveals that the electronic states near the Fermi level are dominated by contributions from $Fe-3d$ and $O-2p$ orbitals, indicating strong hybridization between the transition-metal $d$ states and the oxygen $p$ states. Such hybridization is a characteristic feature of transition-metal oxides with corner-sharing octahedral networks. In the present case, the Fe–O covalency plays a key role in determining both the electronic and structural behavior of the system. The overlap between $Fe-3d$ and $O-2p$ orbitals produces dispersive electronic bands near the Fermi energy and facilitates strong coupling between lattice distortions and electronic states.
The PDOS further shows that finite electronic states cross the Fermi level, indicating that the high-symmetry $P4/mmm$ phase exhibits metallic behavior. This metallic character arises because the undistorted octahedral network does not break the symmetry required to split the $Fe-d$ states or stabilize charge disproportionation. As a result, the Fe ions remain electronically equivalent in this reference structure. 
\subsection{B. Primary structural distortions in the superlattices}
\begin{figure}
\centering
\includegraphics[width=\linewidth]{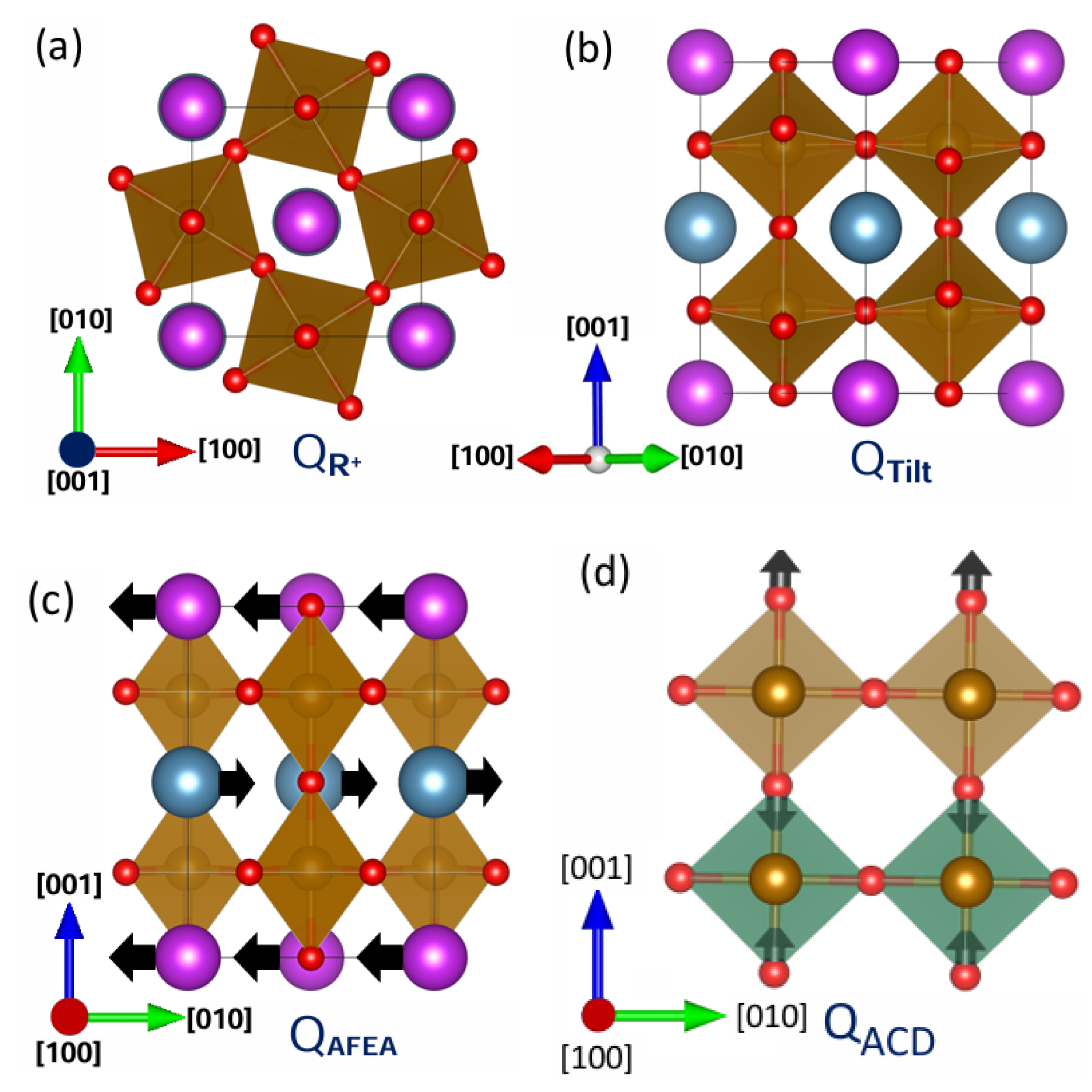}\vspace{-0pt}
\caption {\textbf{Primary structural distortions in BiFeO$_3$/CaFeO$_3$ superlattice that lead to the ground state.} (a) In-phase octahedral rotation Q$_{R+}$ (irrep. $Z_2^+$, $a^0a^0c^+$). (b) Octahedral tilt Q$_{Tilt}$ (irrep. $M_5^-$, $a^-a^-c^0$. (c) Antiferroelectric A-site displacement Q$_{AFEA}$ (irrep. $\Gamma_5^-$). (d) A-type charge disproportionation mode Q$_{ACD}$ (irrep. $\Gamma_3^-$), showing layer-wise alternation of expanded and contracted FeO$_6$ octahedra. The trilinear coupling Q$_{Tri}$ $\sim$ $Q_{R+}Q_{Tilt}Q_{AFEA}$ drives hybrid improper ferroelectricity, which further couples with Q$_{ACD}$ to stabilize the ground state.}
\label{Figure2}
\end{figure}

Figure \ref{Figure2} presents the four fundamental structural distortions that collectively drive the formation of the polar charge-ordered ground state in the BiFeO$_3$/CaFeO$_3$ superlattice. Distortions emerge from the phonon instabilities that establish the ground state in $P4/mmm$ reference structure, including in-phase rotation, tilt, and A-site antiferroelectric displacement, together with the hard A-type charge disproportionation mode. Synergetically, these modes break symmetry and lower the total energy while introducing both polar character and charge ordering. Understanding each distortion's individual symmetry, physical manifestation, and coupling behavior is essential for comprehending how the complex $Pc$ ground state emerges from the cooperative interaction of multiple phonon modes.

The in-phase rotation mode in Figure \ref{Figure2}a, designated $Q_{R+}$ and transforming as the irreducible representation (irrep) $Z_2^+$, represents a cooperative rotation of FeO$_6$ octahedra about the crystallographic $c$-axis. In Glazer's notation, this corresponds to the $a^0a^0c^+$ pattern, indicating that octahedra in adjacent layers along the $c$-direction rotate in the same direction (in-phase) with no rotation about the in-plane axes. This distortion originates from an instability at the Z-point of the Brillouin zone, as revealed by phonon calculations on the high-symmetry $P4/mmm$ phase.

The amplitude of this rotation mode is particularly significant in the BiFeO$_3$/CaFeO$_3$ superlattice because the two constituent materials have different rotational preferences. BiFeO$_3$ in its bulk rhombohedral form exhibits substantial octahedral rotations ($a^-a^-a^-$  in Glazer notation) while CaFeO$_3$ in its low-temperature monoclinic phase displays a complex pattern involving both rotations and breathing distortions. The competition between these tendencies at the interface creates a structural environment where $Q_{R+}$ can develop with considerable amplitude. 

The octahedral tilt mode in Figure \ref{Figure2}b, designated $Q_{Tilt}$ and transforming as the irrep $M_5^-$, describes out-of-phase tilting of FeO$_6$ octahedra about the in-plane axes. In Glazer's notation, this corresponds to the $a^-a^-c^0$ pattern, where octahedra in adjacent layers tilt in opposite directions about both the $a$- and $b$-axes, with no rotation about the $c$-axis. The superscript minus sign indicates the out-of-phase relationship between neighboring octahedra along the tilt direction.

In the superlattice context \cite{rondinelli2011structure}, $Q_{Tilt}$ acquires additional complexity because the tilting pattern must accommodate the different A-site cations (Bi$^{3+}$ and Ca$^{2+}$) in alternating layers. The stereochemically active lone pair on Bi$^{3+}$ creates a local environment that favors specific tilting directions, while the smaller Ca$^{2+}$ cation with no lone pair activity imposes different geometric constraints. This competition between layer-specific structural preferences contributes to the overall amplitude and pattern of the tilt mode. 

The antiferroelectric A-site displacement mode in Figure \ref{Figure2}c, designated $Q_{AFEA}$ and transforming as the irrep $\Gamma_5^-$, involves opposite displacements of Bi and Ca cations within their respective layers. Bi and Ca atoms in successive layers displace in opposite directions along the crystallographic axis. This antiparallel arrangement of unequal cation displacements does produce a microscopic polarization.
The space group resulting from $Q_{AFEA}$ alone is $Pmm2$, which is indeed polar, but the antiparallel pattern of A-site displacements means that the polarization arises primarily from the accompanying oxygen displacements rather than from the cations themselves \cite{benedek2011hybrid}.

The physical origin of this mode lies in the different chemical bonding characteristics of Bi$^{3+}$ and Ca$^{2+}$. Bi$^{3+}$ possesses a stereochemically active $6s^2$ lone-pair that strongly favors off-center displacement within its oxygen coordination polyhedron. This tendency is the same driving force responsible for the large ferroelectric polarization in bulk BiFeO$_3$. Ca$^{2+}$, by contrast, has no lone pair and prefers a more symmetric coordination environment. When forced into the superlattice geometry with alternating Bi-O and Ca-O layers, these competing preferences create a frustrated situation where Bi attempts to displace while Ca resists displacement. The resulting compromise is the antiferroelectric pattern where Bi and Ca move in opposite directions, partially accommodating Bi's displacement tendency while maintaining overall structural coherence.

The A-type charge disproportionation mode in Figure \ref{Figure2}d is a hard mode, designated $Q_{ACD}$ and transforming as the irrep $\Gamma_3^-$, represents the functionally most significant distortion for the electronic properties of the superlattice. This mode describes a breathing-type distortion of the FeO$_6$ octahedra that produces layer-wise alternation of expanded and contracted octahedra. The pattern is termed ``A-type'' by analogy with A-type antiferromagnetic ordering, where ferromagnetic layers are stacked antiferromagnetically. Similarly, in $Q_{ACD}$ all FeO$_6$ octahedra within a given (001) layer exhibit the same breathing distortion (all expanded or all contracted), while adjacent layers alternate between expanded and contracted configurations \cite{shaikh2026polar}.

The physical manifestation of $Q_{ACD}$ involves alternating long and short Fe-O bonds along all directions, but with the important constraint that the pattern is layer-uniform. In an expanded layer, all Fe-O bonds are elongated relative to some average value, while in a contracted layer, all bonds are shortened. This uniform layer-wise breathing creates a charge-ordered state where Fe ions in expanded layers acquire a different oxidation state (formally more reduced, approaching Fe$^{3+}$ with $d^5$ configuration) compared to those in contracted layers (formally more oxidized, approaching Fe$^{4+}$ with $d^4$ configuration). The emergence of $Q_{ACD}$ in the BiFeO$_3$/CaFeO$_3$ superlattice is directly enabled by the interfacial charge imbalance. The superlattice construction places layers that would nominally contain Fe$^{3+}$ (in BiFeO$_3$ blocks) adjacent to layers with nominal Fe$^{4+}$ (in CaFeO$_3$ blocks). This built-in charge modulation provides a charge ordering, and the $Q_{ACD}$ distortion represents the structural relaxation that accommodates and stabilizes this charge pattern.
\begin{figure}
\centering
\includegraphics[width=0.9\linewidth]{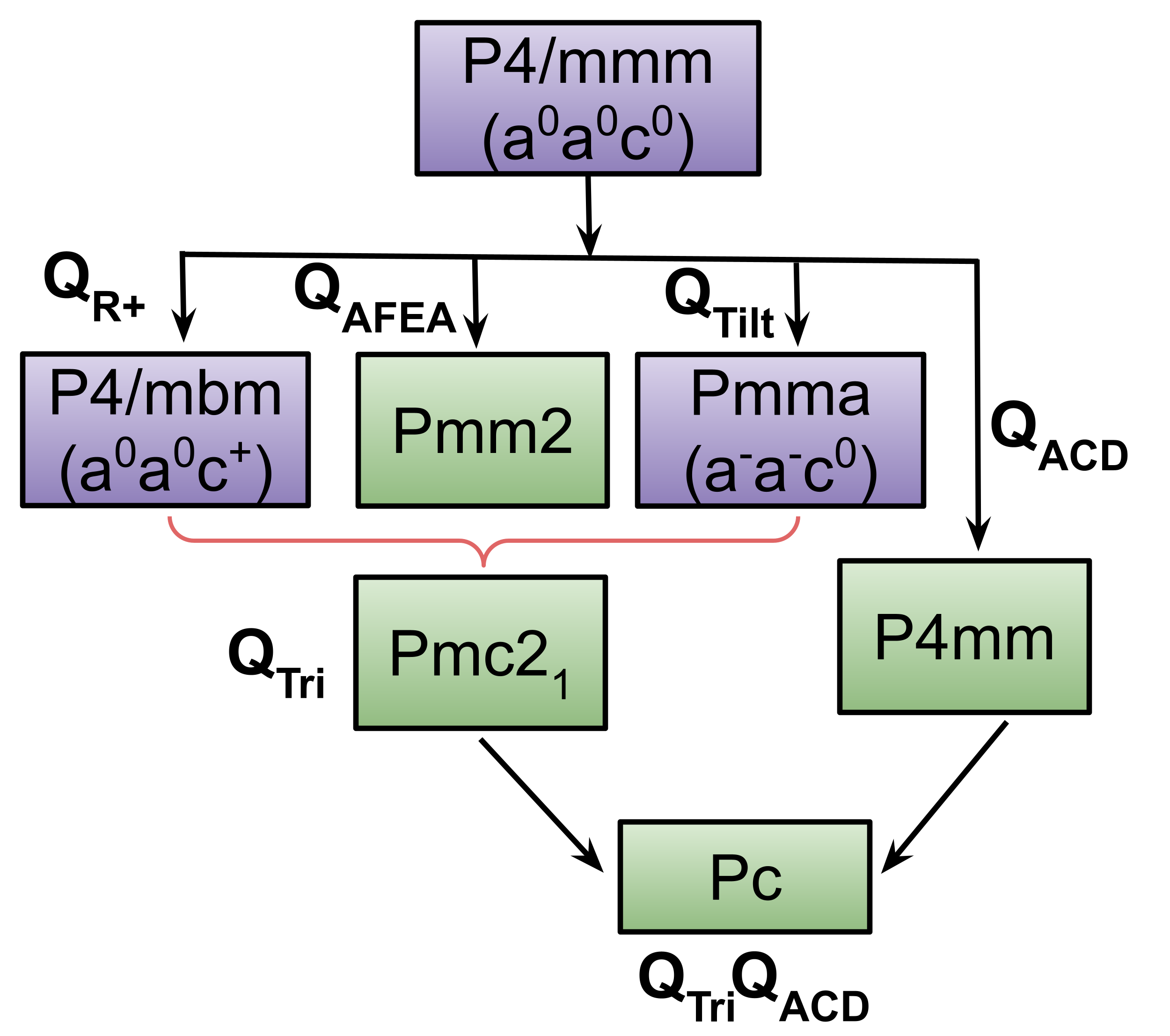}\vspace{-0pt}
\caption {\textbf{Group-subgroup relationship.} The light green color highlights the polar space groups.}
\label{Figure3}
\end{figure}

Figure \ref{Figure3} represents a group-subgroup relation tree that dictates the low symmetry structural ground state \cite{Stokes2016, Campbell2006}, $Pc$-phase, in the BiFeO$_3$/CaFeO$_3$ superlattice. The $Q_{R+}$ mode alone preserves the center of inversion symmetry, as the rotations are geometrically compatible with a centrosymmetric arrangement. When frozen into the undistorted structure, $Q_{R+}$ lowers the symmetry from $P4/mmm$ to the orthorhombic space group $P4/mbm$. This intermediate symmetry represents a common structural motif in many perovskite oxides and serves as a building block for more complex distortion patterns. The $Q_{Tilt}$ mode also preserves the center of inversion symmetry. When frozen into the undistorted structure, $Q_{Tilt}$ lowers the symmetry to $Pmma$. This tilting significantly modifies the orbital overlap between $Fe-3d$ and $O-2p$ states, as the Fe-O-Fe bond angles deviate from the ideal 180$^0$ value found in the undistorted pseudo-cubic structure. The resulting modification of superexchange interactions has profound implications for the magnetic ordering, as the sign and strength of magnetic coupling depend sensitively on these bond angles \cite{Ghosh2015Linear}.

The crucial insight illustrated by Figure \ref{Figure3} is that these four modes do not operate independently. The trilinear coupling term between $Q_{R+}$, $Q_{Tilt}$, and $Q_{AFEA}$ represents the symmetry-allowed interaction. This coupling term is important because it provides a mechanism for hybrid improper ferroelectricity—the generation of a polar state through the cooperative interaction of non-polar structural distortions \cite{benedek2011hybrid, Ghosh2015Linear}. The physics of trilinear coupling can be understood through Landau theory: the free energy of the system contains a term of the form $F \propto \lambda Q_{R+}Q_{Tilt}Q_{AFEA}$ where $\lambda$ is a coupling constant. This term is allowed by symmetry. The presence of this trilinear term means that if two of the modes condense, they create an effective field that drives the condensation of the third mode, even if the third mode is not intrinsically unstable. In the BiFeO$_3$/CaFeO$_3$ superlattice, all three modes are individually unstable as confirmed by phonon calculations (see Supplemental Materials), but their coupling further stabilizes each mode and produces a coherent distortion pattern.

The final step to the ground state involves coupling between $Q_{Tri}$ and the $Q_{ACD}$ mode. This coupling, represented symbolically as $Q_{Tri}Q_{ACD}$, further lowers the symmetry to the monoclinic space group $Pc$ and, crucially, opens the band gap to produce the insulating state. The $Pc$ structure represents the cooperative condensation of all four distortions into a single coherent pattern that simultaneously satisfies the structural ground state.
\subsection{C. Strain-engineered insulator-metal transition in ground state $Pc$-phase}
\begin{figure*}
\centering
\includegraphics[width=\linewidth]{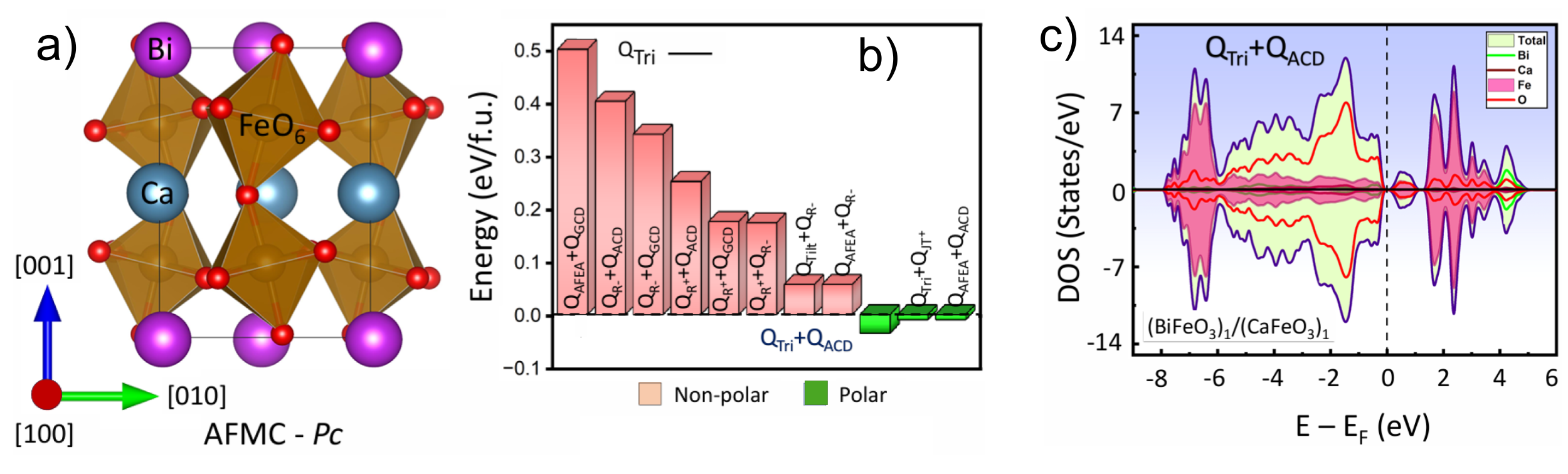}\vspace{-0pt}
\caption {\textbf{Ground state $Pc$ structure of BiFeO$_3$/CaFeO$_3$ superlattice.} (a) Crystal structure with alternating Fe$^{3+}$ (top two octahedra, $3d^5$) and Fe$^{4+}$ (bottom two octahedra, $3d^4$) sites. (b) Relative energies are computed with reference to the Q$_{Tri}$-coupled $Pmc2_1$ phase, and coupled $Q_{Tri}Q_{ACD}$ $Pc$ phase found to be ground state (lower than $Pmc2_1$). (c) Orbital-resolved density of states $Pc$ phase showing strong hybridization between $Fe-3d$ and $O-2p$ orbitals.}
\label{Figure4}
\end{figure*}
Figure \ref{Figure4} presents the ground state $Pc$ structure of the BiFeO$_3$/CaFeO$_3$ superlattice, which emerges from the cooperative coupling between the trilinear mode $Q_{Tri}$ and the A-type charge disproportionation mode $Q_{ACD}$. Panel (a) shows the crystal structure with alternating Fe$^{3+}$ (top two octahedra, $3d^5$) configuration. and Fe$^{4+}$ (bottom two octahedra, $3d^4$) configuration) sites, clearly visualizing the layer-wise charge ordering pattern. This polar charge-ordered arrangement represents the structural manifestation of the $Q_{ACD}$ mode in analogy to A-type antiferromagnetic ordering of the high-symmetry phase, which becomes activated through its coupling with the softer trilinear modes.

Panel (b) provides the energetic hierarchy, showing relative energies computed with reference to the $Q_{Tri}$-coupled $Pmc2_1$ phase. The coupled $Q_{Tri}Q_{ACD}$ structure in space group $Pc$ is found to be lower than $Pmc2_1$, establishing it as the true ground state. The polar, charge-ordered $Pc$ ground state is crucial to understanding its electronic character. We identify that the cooperative coupling between the trilinear mode $Q_{Tri}$ and the A-type charge disproportionation mode $Q_{ACD}$ does not merely lower the structural symmetry but fundamentally alters the electronic behavior, driving a transition from a metal to a semiconductor.

Figure \ref{Figure4}c presents the orbital-resolved density of states (DOS) for this ground state, revealing the mechanism behind this transition. The key feature is the emergence of a distinct energy band gap, confirming the system's transformation into a semiconducting state. An analysis of the PDOS shows that this gap arises from specific and strong hybridization patterns. The top of the valence band is primarily composed of a hybridized mixture of $Fe-3d$ and $O-2p$ states. Critically, we find a dominant role of the $O-2p_z$ and $Fe-3d_z^2$ orbitals in this region. This strong $d-p$ hybridization is pivotal for stabilizing the charge-ordered state.  The bottom of the conduction band is dominated by $Fe-3d$ orbitals, particularly the $d_{xy}$ and $d_{x^2-y^2}$ characters.  The strong hybridization between $Fe-3d_z^2$ and $O-2p_z$ along the $c$-axis (the direction of the superlattice stacking) is instrumental in opening the gap. This specific orbital interaction, stabilized by the collective structural distortions ($Q_{Tri}$ and $Q_{ACD}$), shifts the low-energy $Fe-3d$ orbitals below the Fermi level, resulting in an indirect band gap of about 0.6 eV. This confirms the ground state $Pc$ phase as a polar ferroelectric semiconductor with a substantial spontaneous polarization of 46 $\mu C/cm^2$, a direct consequence of the multimode coupling.

In bulk BiFeO$_3$, ferroelectricity originates from the stereochemically active $6s^2$ lone pair of Bi$^{3+}$, which drives off-center displacements and produces a large spontaneous polarization of $\sim$ 90–100 $\mu C/cm^2$ \cite{Neaton2005PRB}. In the superlattice, Bi atoms are confined to atomically thin layers and are bonded to CaFeO$_3$ layers across interfaces. This confinement disrupts the cooperative lone-pair alignment that exists in the bulk, partially suppressing the polar distortion amplitude. The reduced dimensionality and the need to structurally accommodate the adjacent CaFeO$_3$ layers constrain the Bi off-centering thereby dictating a diminished polarization. Further, charge ordering in CaFeO$_3$ modifies the local crystal field and alters the Fe–O hybridization, which in turn affects the polar distortions. While Bi–O covalency remains important, the Fe–O orbital hybridization (particularly $Fe-3d_z^2$ and$O-2p_z$ along the stacking direction) plays a more dominant role in the superlattice's electronic structure \cite{he2016ferroelectric}. This redistribution of electronic density weakens the effective polarizing field experienced by Bi ions, reducing the overall polarization.
\begin{figure*}
\centering
\includegraphics[width=\linewidth]{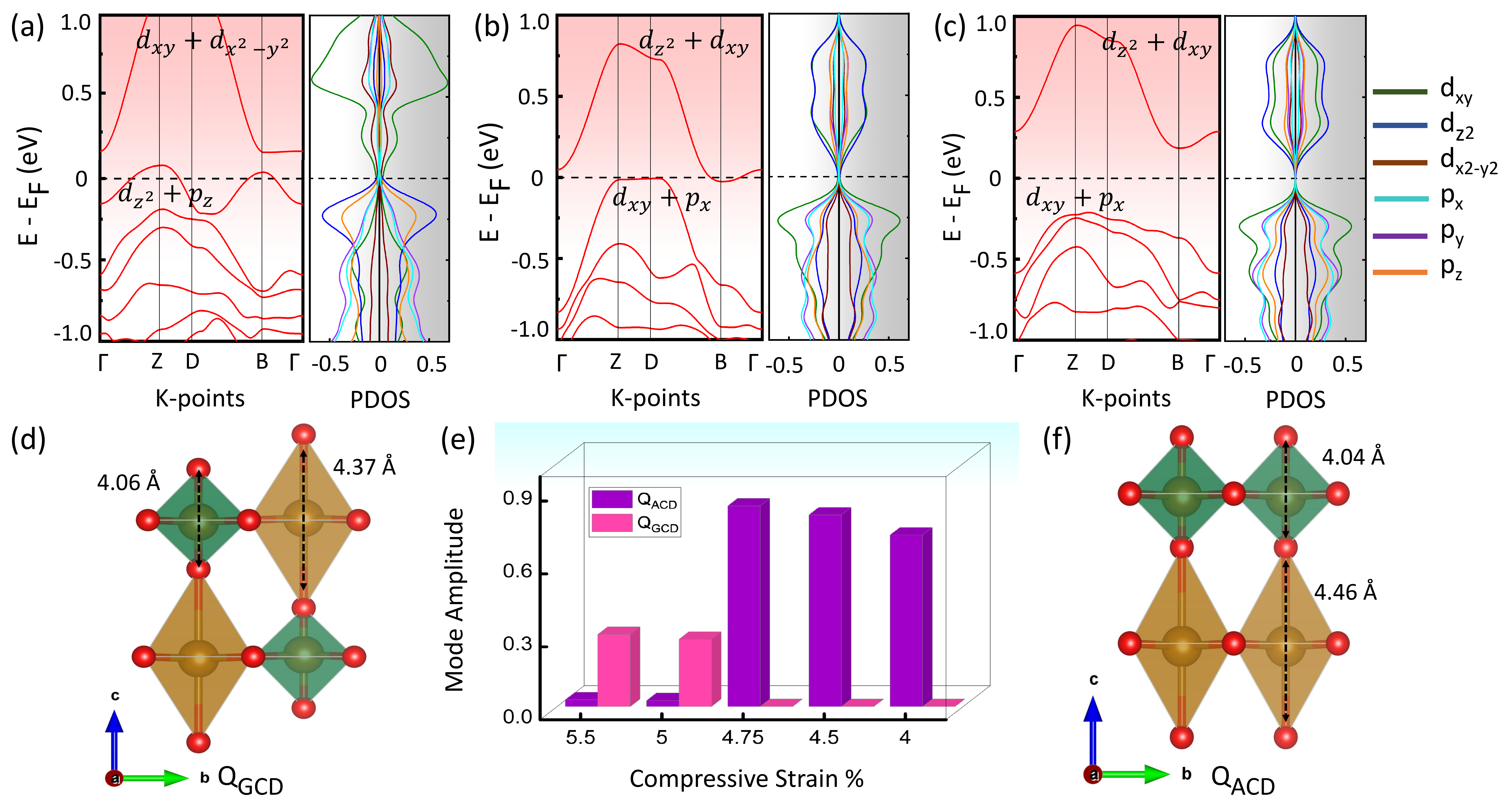}\vspace{-0pt}
\caption {\textbf{Strain-induced metal-insulator transition and charge disproportionation mode crossover in the BiFeO$_3$/CaFeO$_3$ superlattice.} (a-c) Electronic band structure evolution under compressive strain: (a) at $\epsilon$ = -5.0\% (metallic phase), (b) at $\epsilon$ = -4.75\% (critical point with band-touching at the Fermi level, E$_F$), and (c) at $\epsilon$ = -4.5\% (insulating phase with an indirect band gap of $\sim$ 0.3 eV). (e) Amplitude of the two competing charge disproportionation modes as a function of compressive strain: G-type (Q$_{GCD}$, red bars) and A-type (Q$_{ACD}$, purple bars). A clear mode crossover occurs near $\epsilon$ = -4.75\%, coinciding with the metal-insulator transition. (d, f) Schematic representation of the two distinct octahedral breathing patterns: Left --- G-type charge disproportionation (Q$_{GCD}$) showing three-dimensional rock-salt-like alternation of expanded (golden-) and contracted (green-) FeO$_6$ octahedra. Right --- A-type charge disproportionation (Q$_{ACD}$) showing two-dimensional layer-wise alternation of (golden-) and contracted (green-) octahedra, characteristic of the insulating $Pc$ ground state at lower compressive strain. The crossover between these modes, driven by the strong coupling between $Fe-3d$ and $O-2p$ orbitals under strain, enables continuous tuning of the electronic properties from metallic to semiconducting, demonstrating the potential for strain engineering of polar charge-ordered states in ferrite superlattices.}
\label{Figure5}
\end{figure*}

Figure \ref{Figure5} exhibits the central finding of our work. The electronic ground state of the polar $Pc$ phase is not static but can be systematically tuned by an external stimulus. Specifically, it demonstrates that biaxial compressive strain acts as a control knob \cite{Li2025PRL}, driving a continuous mode crossover between two distinct charge disproportionation patterns—Q$_{ACD}$ (Figure \ref{Figure5}d) and Q$_{GCD}$ (Figure \ref{Figure5}f), which in turn governs an insulator-metal transition (IMT) in the ground state. This phenomenon is synergistically linked to the very nature of the superlattice as a platform for strain engineering.

Large compressive strain of several percent has been experimentally realized in BiFeO$_3$-based thin films grown on strongly lattice-mismatched substrates such as LaAlO$_3$ and YAlO$_3$, where compressive strain approaching -4 to -6\% stabilizes highly distorted structural phases in ultrathin epitaxial films. Such strain engineering has been widely demonstrated in oxide heterostructures and superlattices grown by PLD and MBE \cite{zeches2009strain, sando2013crafting, wang2003epitaxial, chakhalian2014Colloquium, hwang2012emergent, mannhart2010oxide, haeni2004room}. In our case, to achieve the compressive strains required to perturb the ground state, one would select a proper substrate. Substrates such as LaAlO$_3$ (LAO) or LSAT, with pseudo-cubic lattice constants that scale to approximately 5.36 $\AA$ and 5.47 $\AA$ in the $\sqrt{2}$ x $\sqrt{2}$ in-plane geometry, respectively, emerge as promising candidates. This coherent epitaxy locks the in-plane lattice parameter of the entire superlattice to that of the substrate. Therefore, the phase diagram presented in Figure \ref{Figure5} is specifically relevant for nanoscale superlattices, where the stored elastic energy is harnessed to create metastable phases, rather than being relieved through plastic deformation. This direct coupling between the macroscopic strain state and the internal atomic coordinates is what enables the exploration of the energy landscape illustrated in the figure.

Figure \ref{Figure5} shows a direct correlation between macroscopic strain, microscopic atomic distortions, and the resulting electronic band structure. The data reveal a tightly coupled mechanism where we observe a specific competition between two symmetry-distinct charge ordering modes. 

\textbf{The Metallic Regime:} Under high compressive strain ($\epsilon$ = -5\%), the octahedral network accommodates the stress by favoring a G-type charge disproportionation (Q$_{GCD}$) pattern. As schematized in Figure \ref{Figure5}d, this mode is characterized by a three-dimensional, rock-salt-like alternation of expanded (Fe$^{3+}$-like, $3d^{5}$) and contracted (Fe$^{4+}$-like, $3d^{4}$) octahedra. Our orbital-resolved analysis shows that this specific distortion pattern stabilizes a metallic state, with significant density of states at the Fermi level arising from a specific orbital ordering ($Fe-3d_z^2$ and $O-2p_z$ at the valence band maxima (VBM); $Fe-3d_{xy}/_{x^2-y^2}$ at the conduction band minima (CBM). In this regime, the strain energy is minimized by a distortion that delocalizes the electronic states.

\textbf{The Critical Point:} As the compressive strain is slightly relaxed ($\epsilon$ $\sim$ -4.75\%), the system approaches a critical point. The band structure shows a characteristic band-touching at the Fermi level, a hallmark of an impending IMT. This is the precise strain state where the lattice can no longer uniquely favor one charge order pattern, leading to maximum electronic instability.

\textbf{The Insulating Regime:} Upon further relaxation to -4.5\% strain, the system crosses a threshold. The A-type charge disproportionation mode (Q$_{ACD}$) becomes the dominant distortion. Figure \ref{Figure5}f illustrates this pattern, showing a two-dimensional, layer-wise alternation of expanded and contracted octahedra. This is the same Q$_{ACD}$ mode that couples with Q$_{Tri}$ to stabilize the ground state $Pc$ structure. Its dominance here re-opens a band gap of $\sim$ 0.3 eV. The orbital characters at the band edges invert, with $Fe-3d_{xy}$ states now populating the VBM, signaling a complete electronic reconstruction driven by the structural crossover.

The synergy is quantitatively captured in Figure \ref{Figure5}e, which plots the amplitudes of the two competing Q$_{CD}$ modes as a function of strain. The graph shows an unambiguous, continuous crossover. The amplitude of the G-type mode (Q$_{GCD}$) decays as strain relaxes, while the A-type mode (Q$_{ACD}$) grows. Their intersection point corresponds precisely to the critical strain ($\epsilon$ $\sim$ -4.75\%) where the band-touching is observed. This demonstrates conclusively that the IMT is not merely an electronic transition but is slave to a structural mode competition.

In short, Figure \ref{Figure5} highlights a complete picture of a strain-engineered phase transition in an oxide superlattice. It shows that by using the tools of modern thin-film growth—specifically, coherent epitaxy on a mismatched substrate—we can access a regime where two fundamental distortive modes (Q$_{ACD}$ and Q$_{GCD}$) compete. The external strain biases this competition. By selecting a substrate that imposes a specific lattice mismatch, one can effectively ``dial in'' a desired electronic phase, from a polar insulator to a correlated metal, and through the critical point in between. This finding positions the BiFeO$_3$/CaFeO$_3$ superlattice as a model system for exploring the interplay of strain, structure, and electronic functionality in complex oxides.
\section{IV. CONCLUSION}
In summary, we have systematically investigated the structural, electronic, and magnetic properties of BiFeO$_3$/CaFeO$_3$ superlattices using first-principles density functional theory combined with symmetry-mode analysis. Our analysis reveals that structural distortions in $P4/mmm$ do not operate independently; rather, their cooperative coupling is essential for stabilizing the ground state. The trilinear coupling $Q_{Tri}\sim Q_{R+}Q_{Tilt}Q_{AFEA}$ through hybrid improper ferroelectricity—generates a polar $Pmc2_1$ structure, which further couples with the charge disproportionation mode $Q_{ACD}$ to produce a lower-symmetry monoclinic $Pc$ phase. This final structure, lying below the $Pmc2_1$ phase, represents the ground state of the superlattice.

The cooperative condensation of these structural distortions fundamentally alters the electronic behavior of the system. While $P4/mmm$ phase exhibits metallic character arising from electronically equivalent Fe sites, the polar $Pc$ ground state transforms into a ferroelectric semiconductor with an indirect band gap of around 0.6 eV and a substantial spontaneous polarization of 46 $\mu C/cm^2$. This metal-to-insulator transition originates from the activation of $Q_{ACD}$, which produces layer-wise alternation of expanded (Fe$^{3+}$-like, $3d^{5}$) and contracted (Fe$^{4+}$-like, $3d^{4}$) octahedra. The resulting charge-ordered state is accompanied by strong hybridization between $Fe-3d_z^2$ and $O-2p_z$ orbitals along the stacking direction, which shifts the low-energy $Fe-3d$ states below the Fermi level and opens the band gap. The ground state exhibits C-type antiferromagnetic ordering, demonstrating the intimate coupling between structural distortions, charge ordering, and magnetic order in this system.

A particularly significant finding of our work is the remarkable tunability of the ground state electronic properties under external strain. By exploiting the artificial superlattice geometry, which permits coherent epitaxial growth on lattice-mismatched substrates, we demonstrate that biaxial compressive strain can drive a continuous mode crossover between two distinct charge disproportionation patterns. Under high compressive strain ($\epsilon$ = -5.0\%), the system favors a G-type charge disproportionation ($Q_{GCD}$) characterized by three-dimensional rock-salt-like alternation of expanded and contracted octahedra, stabilizing a metallic state. As strain relaxes to $\epsilon$ = -4.5\%, the A-type mode ($Q_{ACD}$) becomes dominant, restoring the polar insulating phase. At the critical strain $\epsilon$ $\sim$ -4.75\%, a characteristic band-touching phenomenon at the Fermi level was observed. This strain-controlled mode crossover, quantitatively captured through amplitude analysis of the competing distortions, demonstrates that the insulator-metal transition is slave to a structural competition rather than a purely electronic effect.

Our results identify BiFeO$_3$/CaFeO$_3$ superlattice as a model system for exploring the interplay between strain, structural distortions, charge ordering, and electronic functionality in complex oxide heterostructures. The ability to continuously tune the electronic properties—from polar insulator to correlated metal and through the critical point in between—by selecting appropriate substrates (such as LAO or LSAT) opens new avenues for designing functional materials with applications in next-generation electronic, spintronic, and multiferroic devices. Further, this work demonstrates that the cooperative coupling of multiple structural instabilities, guided by symmetry principles and controlled through interface engineering, provides an effective strategy for discovering and designing emergent electronic phases in artificial oxide superlattices.
\section{ACKNOWLEDGEMENT}
\par
R.G acknowledges SRMIST for her fellowship. M.S. and S. G. acknowledge support from the Center for Advanced Computational and Theoretical Sciences and the Department of Physics and Nanotechnology, SRM Institute of Science and Technology, Kattankulathur-603 203, Tamil Nadu, India.  Computations were performed using the High Performance Computing Cluster at SRM Institute of Science and Technology, Chennai, India. 
%

\end{document}